\documentclass[lettersize,journal]{IEEEtran}
\usepackage{amsmath,amsfonts}
\usepackage{algorithmic}
\usepackage{algorithm}
\usepackage{array}
\usepackage[caption=false,font=normalsize,labelfont=sf,textfont=sf]{subfig}
\usepackage{textcomp}
\usepackage{stfloats}
\usepackage{url}
\usepackage{verbatim}
\usepackage{bm}
\usepackage{graphicx}
\usepackage{svg}
\usepackage{cite}
\usepackage{tabularx,booktabs}
\usepackage[table]{xcolor}
\usepackage{makecell, multirow}
\usepackage{hhline}
\usepackage[hidelinks]{hyperref}
\usepackage{soul}
\begin{document}

\title{An Annual Quasi-Static Time-Series Simulation Framework for Enhanced Transmission System Expansion Planning}

\author{Hussein Suprême, 
        Martin de Montigny,
        Kevin-R. Sorto-Ventura, 
        Hind Chit Dirani, 
        Mouhamadou Makhtar Dione, 
        Nicolas Compas
\thanks{Manuscript received April 3, 2026.}
\thanks{Hussein Suprême, Martin de Montigny, Kevin-R. Sorto-Ventura, Hind Chit Dirani and Mouhamadou Makhtar Dione are with Institut de recherche d’Hydro-Québec, Varennes, QC, Canada, J3X 1S1. (e-mail: \{supreme.hussein2, demontigny.martin, sorto-ventura.kevin-rafael, chitdirani.hind2, dione.mouhamadoumakhtar\}@hydroquebec.com).
Nicolas Compas is with Hydro-Québec, Montréal, QC, Canada, H2Z 1A4. (e-mail: compas.nicolas@hydroquebec.com).}%
}



\maketitle

\begin{abstract}
The increasing integration of distributed energy resources (DERs), variable renewable energy sources, and emerging technologies presents new challenges for transmission system expansion planning (TSEP). Traditional snapshot-based and deterministic approaches are inadequate for capturing the temporal dynamics and operational constraints of modern power systems. This paper introduces an annual quasi-static time-series simulation (AQSTSS) framework that enables high-resolution, year-round modeling of transmission systems, incorporating detailed equipment behavior, control strategies, and DER interactions. By simulating system performance across all seasons and operating conditions, AQSTSS uncovers flexibility opportunities and operational constraints that static methods overlook. Applied to Hydro-Québec’s projected 2035/2036 grid, the framework reveals critical insights under high wind and electric vehicle penetration. It also integrates an energy storage control strategy designed to mitigate wind variability and support grid reliability. Furthermore, AQSTSS facilitates the assessment of system resilience under diverse scenarios, including extreme weather and load variability. The simulation results underscore the importance of aligning planning with operational realities to ensure secure, efficient, and future-ready grid development. Overall, the proposed framework enhances the robustness of TSEP by bridging the gap between long-term planning and real-time operational needs.
\end{abstract}


\section{Introduction}\label{sec:intro}
\IEEEPARstart{T}{o} achieve a smooth energy transition and reach net zero emissions by 2050, future power systems must evolve across five key dimensions (5D): decentralization, decarbonization, digitalization, deregulation, and democratization~\cite{Moghaddam2022}. The first two dimensions significantly increased the penetration of variable renewable energy sources into transmission systems (TS). In addition, distribution systems are transforming into active components owing to the high penetration of distributed energy resources (DERs). Digitalization, while enabling advanced monitoring and control, has also expanded the attack surface to cyber threats. Simultaneously, climate change is driving more frequent and severe extreme weather events, further stressing grid resilience. Deregulation and democratization amplify competition and uncertainties in the planning and operation of power systems. This paradigm shift presents both market opportunities and operational challenges for transmission system planners (TSPs) and operators.

Recent studies, primarily focusing on distribution system planning, have emphasized the limitations of static models and highlighted the need for time-series-based simulation frameworks capable of capturing the operational dynamics of DERs), energy storage systems (ESSs), and control mechanisms over extended horizon~\cite{Deboever2018}. For instance, hybrid clustering and time-series aggregation techniques have been proposed to preserve both chronological consistency and extreme event characteristics in long-term planning models~\cite{Novo2022, Moradi2023}. Moreover, reliability-constrained planning models integrate probabilistic forecasting of load and renewable generation to improve the accuracy of net demand projections~\cite{Choi2021}. The integration of DERs and ESSs into distribution system planning has also evolved, with control-aware frameworks incorporating model predictive control, distributed optimization, and DER management systems~\cite{Sarajpoor2021}. Building on these recent developments, transmission system planning methodologies must evolve toward an integrated approach that accounts for technological transformations driven by energy transition.

{Traditional transmission system expansion planning (TSEP) methods are insufficient for addressing the variability and uncertainties introduced by emerging energy technologies while maintaining reliable system operation. Existing approaches are generally classified as deterministic, focused 
on worst‑case or representative operating 
points; and 
nondeterministic, which explicitly incorporate uncertainty through multiple operating conditions~\cite{Li2011}. Most TSEP formulations rely on simplified mathematical models, commonly using DC power flow calculations (PFCs), and rarely account for system control actions~\cite{Lumbreras2016}. To better capture uncertainty, probabilistic power flow (PPF) and scenario‑based techniques (SBT) have been used~\cite{Bugyi2003}. PPF derives probability distributions of system variables (e.g., nodal voltages, branch flows) when loads and generation are modeled as random variables, while SBT generates multiple scenarios representing possible future system states or relies on probabilistic models to synthesize such scenarios. However, these approaches typically treat system states as time‑independent and do not capture temporal correlations.}

To address this gap, we propose an Annual Quasi‑Static Time‑Series Simulation (AQSTSS) framework for TSEP. The quasi-static time-series (QSTS) simulation consists of sequential steady‑state AC PFCs, where each solved state becomes the initial condition for the next. Extending this process to a full‑year horizon, the AQSTSS framework produces chronology‑preserving trajectories that capture load diversity, renewable variability, and device‑level control actions. Using a 5‑minute resolution, 
it models tap‑changer operations, storage dispatch, and demand‑side management (DSM). This enables planners to assess the critical off‑peak operating states that increasingly govern constraints under high variable renewable energy (VRE) penetration, providing detailed yearlong insights into losses, curtailment, ramping behavior, discrete control actions, and equipment outages.
\color{black}
\begin{figure}[htbp]
    \centering
    \includegraphics[width=1.0\columnwidth]{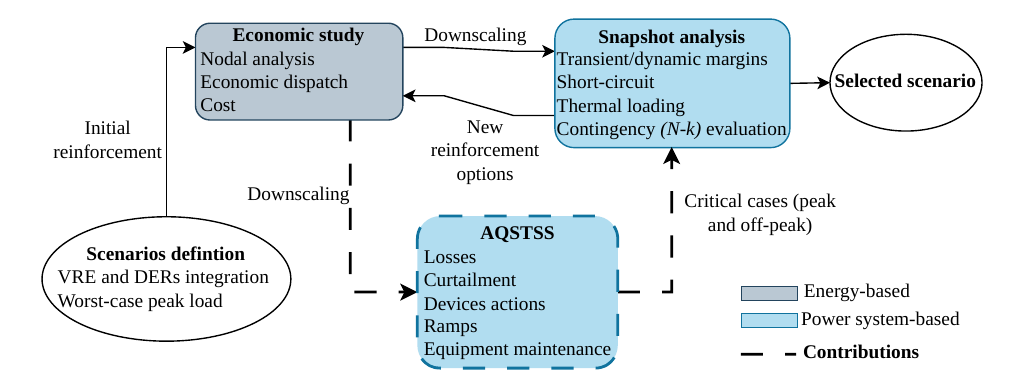}
    \caption{Integration of the AQSTSS layer within the traditional TSEP process}
    \label{fig:contribution}
\end{figure}

{
As shown in Fig.~\ref{fig:contribution}, the AQSTSS framework integrates seamlessly into the conventional TSEP process, providing the temporal resolution absent from energy‑based economic studies and steady‑state technical assessments. By translating long‑term scenarios into chronological operating states and exposing dimensioning off‑peak conditions, AQSTSS reveals reinforcement needs overlooked by snapshot analyses, thereby supporting a more reliable and control‑aware planning process.
}
The main contributions of this paper are as follows:
\begin{enumerate}
     \item A full‑year, chronologically consistent QSTS simulation framework is developed, providing planners with time‑dependent network conditions that directly reveal where, when and why reinforcements are needed.
    \item An energy storage control strategy is proposed to mitigate the seasonal variability of wind generation, leveraging the chronological information. 
    \item An integrated analysis methodology is demonstrated, combining annual operational outcomes with AQSTSS results to validate the applicability and effectiveness of the proposed approach for long‑term TSEP.
\end{enumerate}
\color{black}

The remainder of this paper is organized as follows. Section~\ref{sec:aqstss} outlines the inputs and requirements for AQSTSS across applications, such as power quality, operational deviations, equipment utilization, and DER impact. Section~\ref{sec:ess_strategy} presents the proposed energy storage-based control strategy. Sections~\ref{sec:case_studies} and~\ref{sec:results} describe case studies and discuss the corresponding results. Finally, Section~\ref{sec:conclusion} concludes the paper.

\section{Proposed AQSTSS Approach}\label{sec:aqstss}
\begin{figure*}[htbp]
    \centering
    \includegraphics[width=\textwidth]{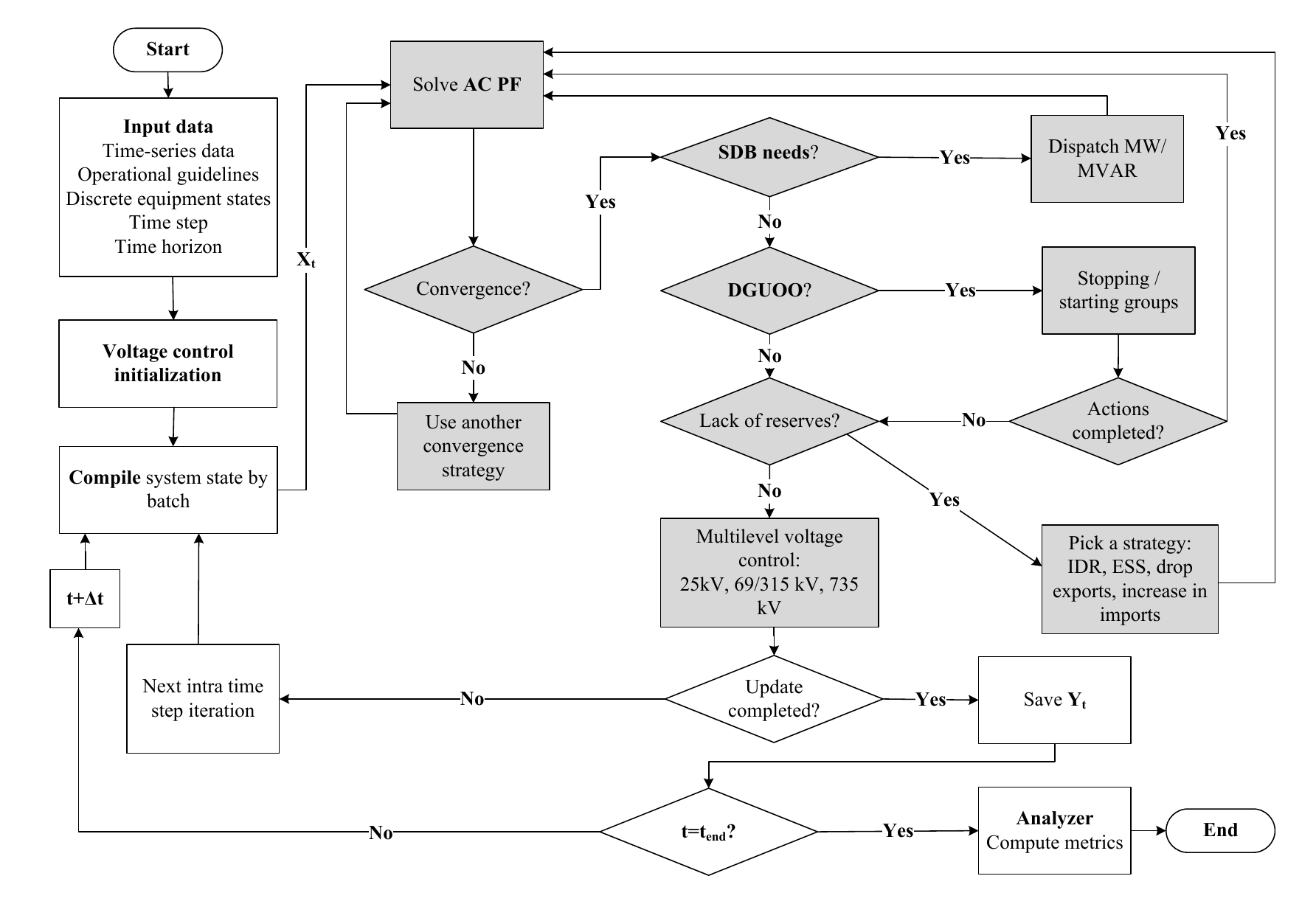}
    \caption{Overview of the AQSTSS algorithm}
    \label{fig:overview_algo}
\end{figure*}

Snapshot‑based analyses, which examine only a few isolated operating conditions, often provide overly conservative or incomplete assessments of system performance when integrating DERs and emerging technologies. Prior work~\cite{Supreme2024} highlighted the value of QSTS simulations for transmission planning but was limited to three consecutive days near the peak planned system (PPS), preventing a full characterization of temporal variability. The proposed AQSTSS framework, shown in Fig.~\ref{fig:overview_algo}, extends this approach to a full‑year horizon, enabling a comprehensive and temporally resolved assessment of system operability, congestion patterns, and flexibility needs. This section outlines the principal methodological components underlying the proposed approach. 
\color{black}
\vspace{-11pt}
\subsection{Data Inputs and Output Responses}\label{subsec:io}
The proposed AQSTSS adopts a rule‑based methodology that embeds the system’s spatial and temporal characteristics within a unified simulation architecture. The main core is the virtual operator (VO), whose decision processes and corrective actions govern the simulation workflow and are depicted by the greyed blocks in Fig.~\ref{fig:overview_algo}. 
The framework relies on four input classes: \textbf{i)} a detailed future TS model; \textbf{ii)} the network’s operational guidelines and constraints; \textbf{iii)} synchronized time‑series forecasts for loads, generation, and interconnections; and \textbf{iv)} the technical characteristics of TS‑coupled resources. Operational security thresholds derived from the guidelines serve as decision criteria for the VO. Furthermore, DSM resources, such as interruptible demand response (IDR) and ESSs, are modeled through parameters including activation time, duration, and operating limits, and enhance system flexibility by supporting supply–demand balance (SDB) and controlling power flows through critical assets. These inputs collectively define the system state vector $\bm{X_t}$ at each time-step $\bm{t}$.

The simulation starts from a planned network configuration, typically corresponding to peak‑load conditions, with all transmission system components represented up to the distribution interfaces across 25–735 kV voltage levels. To incorporate operational realism, the model embeds key steady‑state constraints, including thermal limits, reserve requirements, shunt operating ranges, and voltage boundaries. As illustrated in  Fig.~\ref{fig:overview_algo}, after each time-step, the framework updates the network to a new converged steady‑state solution, which then serves as the initial condition for the next step to ensure temporal consistency. Device states that evolve over time, such as transformer tap positions, are tracked explicitly, and the resulting system state is stored in the matrix $\bm{Y_t}$.
\color{black}

The analyzer plays a key role in processing simulation outputs and evaluating a comprehensive set of both continuous and discrete performance metrics. It quantifies transformer tap‑changer operations, capacitor and reactor switching actions, the availability of reactive power for voltage support, active power losses, and system reserve levels. By synthesizing these indicators, the analyzer enables a detailed and holistic assessment of system performance, operational efficiency, and overall grid reliability.
\subsection{System Monitoring and Resolution Techniques}\label{subsec:resolution}
The VO maintains physical coherence, operational security, and numerical robustness by continuously monitoring SDB, reserves, generator performance, and voltages throughout the simulation horizon. Transitions between peak and low‑load periods remain particularly challenging, as they require tight coordination of controllable assets to preserve feasibility. The VO manages these operating shifts to ensure power‑flow convergence and reliable system operability at every time-step.
\color{black}

\vspace{-9pt}
The active power $P_i(t)$ and $P_i(t+1)$ in~\eqref{eq:Pi_t+1} are time-dependent,
capturing variations in DERs, load, and generation. Similarly, as shown in \eqref{eq:Qi_t+1}, the reactive power $Q_i(t)$ evolves to reflect the changes in both conventional loads and DER contributions. 
As depicted in Fig.~\ref{fig:overview_algo}, the proposed framework resolves each time-step through a series of incremental updates, during which the VO ensures that all intermediate states remain feasible and operationally valid. This contrasts sharply with traditional QSTS formulations, where large, full-step injections are applied prior to solving the AC power flow. Such full-step updates are known to introduce non‑physical energy accumulation at the swing bus, often leading to convergence failures and unstable behavior in PFCs. By embedding feasibility checks within each incremental update, 
the method offers a more reliable alternative that mitigates numerical pathologies and maintains controller consistency throughout the simulation horizon. 
\color{black}
At each intra-time step iteration $j$, 
the AC PFCs in \eqref{eq:Pi_t}-\eqref{eq:Qi_t} are solved using the Newton-Raphson method to update the voltage magnitudes and angles, as shown in \eqref{eq:Vi}-\eqref{eq:deli}. 
This power‑flow solution also identifies and diagnoses any security violations. 
\color{black}
If the solver diverges, fallback strategies are applied, including tuning Newton–Raphson parameters, switching to its decoupled form, and adjusting shunt‑VAR constraints~\cite{Supreme2025}.
\begin{equation}\label{eq:Pi_t+1}
P_i(t+1) = P_i(t) + \sum_{j=1}^{J} \Delta P_i^j(t)
\end{equation}
\begin{equation}\label{eq:Qi_t+1}
Q_i(t+1) = Q_i(t) + \sum_{j=1}^{J} \Delta Q_i^j(t) + \Delta Q_{i,DER}^j(t)
\end{equation}
\begin{align}\label{eq:Pi_t}
P_i^j(t) = \left|V_i^j(t)\right| \sum_{k=1}^{N}\left|V_k^j(t)\right| &\left(G_{ik}^j \cos{\delta_{ik}^j(t)} \right. \nonumber \\
&\left. + B_{ik}^j \sin{\delta_{ik}^j(t)}\right)
\end{align}
\begin{align}\label{eq:Qi_t}
Q_i^j(t) = \left|V_i^j(t)\right| \sum_{k=1}^{N}\left|V_k^j(t)\right| &\left(G_{ik}^j \sin{\delta_{ik}^j(t)} \right. \nonumber \\
&\left. - B_{ik}^j \cos{\delta_{ik}^j(t)}\right)
\end{align}
\begin{equation}\label{eq:Vi}
\left|V_i^{j+1}(t)\right| =\left|V_i^{j}(t)\right| + \Delta \left|V_i^{j}(t)\right|
\end{equation}
\begin{equation}\label{eq:deli}
\delta_{i}^{j+1}(t) = \delta_{i}^j(t) + \Delta \delta_{i}^j(t)
\end{equation}

Both active and reactive power outputs are adjusted to restore the SDB, subject to generator ramp‑rate limits. The balancing action is allocated proportionally across participating units according to their available upward or downward regulation margins, depending on the direction of the imbalance detected at the swing bus. The decision process then continues with the dynamic coordination of automatic generation control (AGC) units, guided by the deviation of generating units from optimal operation (DGUOO), defined as the difference between each regulated unit’s current output and its optimal dispatch point.
Generators start-ups or shutdowns are triggered to maintain this indicator within its predefined safety bounds. \color{black} 
Subsequently, the system ensures adequate 
AGC reserve margin, which is computed by summing the difference between maximum power and current power of all generating units participating in AGC. \color{black}
If the reserve margin falls below the minimum level specified in the operational directive, a hierarchical sequence of corrective actions is initiated. 
These include additional generator start-ups, controlled voltage reductions, demand management actions, and adjustments of import/export schedules. Each corrective action requires a new power‑flow solve to maintain system security. 
\color{black}

\begin{figure}[htbp]
    \centering
    \includegraphics[width=1.0\columnwidth]{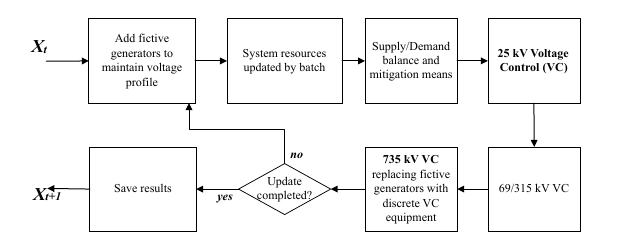}
    \caption{Overview of voltage control algorithms from 25 to 735 kV}
    \label{fig:overview_vctrl}
\end{figure}

Once the constraints related to SDB, DGUOO, and reserves are satisfied, \color{black}
a coordinated multi-stage voltage control technique, presented in~\cite{Supreme2025}, is applied iteratively using temporary fictitious generators to maintain the target voltage profiles (see Fig.~\ref{fig:overview_vctrl}). Control begins at the 25 kV level through shunt component operation, based on availability and regional thresholds, and transformer tap adjustments. At the 735 kV level, control actions include line reconnection to enhance power transfer capability, shunt switching for voltage stabilization, line disconnection to maintain the voltage within secure operational limits while meeting power transfer requirements, and compensator setpoint adjustments for reactive power control. 
The removal of fictitious generators and the full discretization of all reactive power–producing equipment ensure that network voltages remain physically coherent and consistent with actual device capabilities. Once voltage feasibility is restored, control returns to the resource‑update mechanism to process the next batch of operational adjustments. When all resources have been updated, the framework checks for convergence: if a feasible, constraint‑satisfying solution is obtained, the operating point is recorded for the current time-step; otherwise, internal iterations continue until the system reaches a valid solution.
\color{black}
\subsection{Temporal Parallelization Resolution and Initialization}\label{subsec:initialization}
A temporal parallelization strategy is implemented in the proposed AQSTSS framework to efficiently manage the computational demands of year‑round simulations. The annual horizon is divided into weekly intervals, each of which is executed independently on a dedicated processor within a computing cluster orchestrated by the HTCondor software.  
This design provides both scalability and enhanced temporal resolution while significantly reducing total computation time. Continuity across weeks is inherently preserved because all operational and resource profiles remain fully consistent across the divided segments. As a result, any residual discontinuities at weekly boundaries remain negligible.
\color{black}

To initialize the simulations, static and synchronous compensator setpoints are adjusted to meet the target voltage levels and control deadbands, ensuring that the system begins from a stable and realistic operating point. A preliminary power flow analysis is then performed to verify the system stability and identify any voltage deviations or imbalances introduced by the scheduled load and generation profiles. The simulation proceeds iteratively with continuous refinement of the control parameters and logic to improve the voltage regulation. Throughout the process, active power flows are monitored to maintain a dynamic balance between generation and demand, particularly under variable operating conditions driven by DERs and load fluctuations. This integrated approach ensures efficient system operation and compliance with predefined performance and reliability criteria.
\subsection{Simulation Time-Step Resolution}\label{subsec:time_step}
The time‑step resolution and simulation horizon in AQSTSS are selected according to the study objectives and the operational characteristics of the technologies under analysis. A 5‑minute resolution is adopted for impact assessments involving ESSs~\cite{Sorto-Ventura2024}, electric vehicles, and distributed photovoltaics (DPVs)~\cite{Supreme2024}, as it captures the cyclic charging and discharging of storage, the variability of DPV injections, and the evolution of steady‑state voltages. This finer granularity is also required to accurately represent the operation of discrete control devices—including shunt and line switching, transformer tap changes, and generator commitment actions—as well as to evaluate ramping behavior, operational reserves, and SDB security in applications such as import/export management and demand response. 
In contrast, an hourly time-step is sufficient for studies focused on long‑term energy trends, 
where short‑term dynamics play a limited role.

In general, finer time-steps should be used when system behavior is driven by fast‑acting resources, discrete control operations, or short‑term security constraints, whereas coarser resolutions are adequate when the objective is to characterize long‑term energy patterns or aggregate power flows. This guideline provides a consistent basis for selecting an appropriate resolution independent of the specific case under study.
\color{black}
\section{Energy Storage-Based Control Strategies for Reducing Wind Generation Variability}\label{sec:ess_strategy}
Previous works have addressed wind power variability through various energy management systems and ESS power dispatch algorithms, often aiming to balance technical and economic objectives in renewable generation \cite{Yang2022, Yang2023}. For instance, \cite{Jannati2020} proposed a battery energy storage system (BESS) power allocation strategy for wind power smoothing that also extends battery life. 
Similarly,~\cite{Li2013} demonstrated how multiple BESSs can be coordinated to smooth renewable generation while regulating state-of-charge (SOC) to limit degradation.

{This section presents an ESS control strategy aimed at dispatching power to mitigate wind generation variability while regulating the SOC. This strategy is adapted from~\cite{Jannati2020, Li2013} and is implemented in the proposed AQSTSS framework to enable TSEP studies that involve both wind farms and ESSs. The threshold-based control method, presented in  Algorithm~\ref{alg:ess_mode}, is applied to the generic ESS model for QSTS simulations developed in~\cite{Sorto-Ventura2024}.}
\begin{algorithm}
	\caption{Operational Modes of ESS for Mitigating \\Production Variability in Wind Farms}
	\label{alg:ess_mode}
	\begin{algorithmic}[1]
    	\renewcommand{\algorithmicrequire}{\textbf{Inputs:}}
    	\renewcommand{\algorithmicensure}{\textbf{Output:}}
    	\REQUIRE $t_{start}^{peak}, t_{end}^{peak}, Gen_{t}^i, Gen_{max,lim}^i, Gen_{min,lim}^i$
    	\ENSURE  \textit{Mode of operation}
    	\\ \textit{Start} :
        \\ \text{\# Surplus generation from wind farms}
    	 \IF {$Gen_{t}^i \geq Gen_{max,lim}^i$}
    	 	\IF {$t_{start}^{peak} < t < t_{end}^{peak}$}
    			 \STATE \textit{Mode of operation}: \textbf{Standby}
                \ELSE
    	 	 \STATE \textit{Mode of operation}: \textbf{Charging}
    	      \ENDIF
        \\ \text{\# Shortage of generation from wind farms}
    	 \ELSIF {$Gen_{t}^i \leq Gen_{min,lim}^i$}
    	 	\STATE \textit{Mode of operation}: \textbf{Discharging}
        \\ \text{\# Generation within limits}
    	 \ELSIF{$Gen_{min,lim}^i < Gen_{t}^i < Gen_{max,lim}^i$}
        	 	\IF {$SOC_t < SOC_{bal}$}
        			 \STATE \textit{Mode of operation}: \textbf{Charging}
                    \ELSIF{$SOC_t > SOC_{bal}$}
        	 	 \STATE \textit{Mode of operation}: \textbf{Discharging}
        	      \ENDIF
          \ELSE
            \STATE \textit{Mode of operation}: \textbf{Standby}
    	 \ENDIF
    	\RETURN \textit{Mode of operation} 
	\end{algorithmic} 
\end{algorithm}


According to Algorithm \ref{alg:ess_mode}, the ESS discharges or charges to maintain the total wind generation within the established upper and lower generation limits $Gen_{max,lim}^i$ and $Gen_{min,lim}^i$. Moreover, the ESS is prevented from charging during peak-load periods, denoted $[t_{start}^{peak}, t_{end}^{peak}]$, as the power availability is limited. When the wind power $Gen_t^i$ lies within the generation limits, the ESS discharges or charges until it reaches the balance energy level $SOC_{bal}$.
{Finally, it should be noted that the control strategy is executed at every time-step of the simulation without any knowledge of future wind generation values. Consequently, upper and lower generation limits along with the SOC balance level are parameters, set by the user, which are independent of the wind generation profiles.}

\section{Case Studies}\label{sec:case_studies}
The Hydro-Québec transmission system projected for 2035/2036 was modeled with all its equipment from 25 kV to 735 kV, including their discrete states. It connects large hydroelectric plants in the north to load centers in the south via thousands of kilometers of transmission lines \cite{Supreme2022}.

To highlight the relevance of the proposed AQSTSS framework, two scenarios were analyzed. As summarized in Table~\ref{tab:cases}, \textbf{Scenario 1} (\textbf{S1}) assumes a high penetration of EVs, with 1.7 million units representing approximately 2.5 GW of additional load, along with 6 GW of the new wind generation capacity. These wind farms are geographically distributed across two regions, denoted by $i \in\{\text{East, non-East}\}$. To capture realistic wind variability, the generation profiles of the new wind farms were derived from the historical data of neighboring installations and segmented into five seasonal periods, $p \in\{1,2,3,4,5\}$, reflecting distinct operating conditions throughout the year.
\begin{table}[htbp]
    \caption{Transmission System Evolution with Projected DERs and Wind Energy Penetration}
    \centering
        \begin{tabular}{c c c c}
            {}&\makecell{\textbf{Current TS}\\ \textbf{(2024)}} & \makecell{\textbf{S1}\\ \textbf{(2032)}} & \makecell{\textbf{S2}\\ \textbf{(2035)}}\\
            \hline
            \rowcolor{blue!10}
            \textbf{Wind power} & $\leq$ 4 GW & $\approx$ 10 GW & $\approx$ 10 GW\\
            \textbf{EVs} & $\leq$ 0.25 MW & 2.5 GW & 2.5 GW\\
            \rowcolor{blue!10}
            \textbf{ESSs} & Negligible & Negligible & 1.5 GW\\
            \hline
        \end{tabular}
        \label{tab:cases}
\end{table}

\textbf{Scenario 2} (\textbf{S2}) introduces two ESS aggregations with a combined power rating of 1.5 GW into the network to offset the wind generation variability. As shown in \eqref{eq:gen-max} and \eqref{eq:gen-min}, the maximum and minimum generation limits of each division were calculated based on the mean $\mu_p^i$ and standard deviation $\sigma_p^i$ of the profiles per period. This statistical approach ensures that the ESS dispatch is aligned with the expected variability and uncertainty of the renewable generation.
\begin{equation}\label{eq:gen-max}
Gen_{max,lim}^i = \mu_p^i + 1.5\times \sigma_p^i
\end{equation}
\begin{equation}\label{eq:gen-min}
Gen_{min,lim}^i = \mu_p^i - 1.5\times \sigma_p^i
\end{equation}

\section{Results and Discussion}\label{sec:results}
This section presents the types of results generated using the proposed AQSTSS framework, which can support transmission planners in evaluating the impacts of energy transition and identifying non-wire alternatives for integrated transmission system planning. For each scenario, tens of millions of power-flow calculations were executed through internal iterations to ensure that the resulting network states were both secure and operable under a wide range of conditions.

\subsection{Annual Load Distribution}\label{subsec:load_distribution}
\begin{figure}[htbp]
    \centering
    \includegraphics[width=1.0\columnwidth]{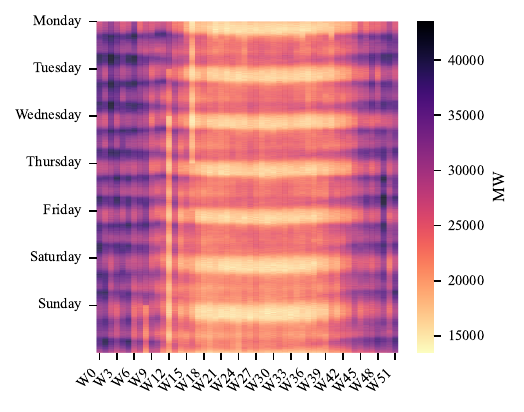}
    \caption{Total load demand distribution}
    \label{fig:heatmap}
\end{figure}

Fig.~\ref{fig:heatmap} illustrates the distribution of the total annual demand in \textbf{S1}, categorized by day and week, revealing pronounced seasonal variations in the load profiles. These fluctuations underscore the need for adaptive operational strategies and robust infrastructure to maintain grid reliability throughout the year. During winter peak periods (weeks W1–W3), the load demand exceeds 40 GW, requiring extensive use of transmission infrastructure to supply major load centers, along with the widespread use of shunt devices to maintain voltage levels and ensure system security. In contrast, summer troughs (weeks W31–W35) can drop as low as 13 GW, requiring the timely disconnection of shunt capacitors and lines and the connection of shunt reactors to prevent overvoltage. These actions must be strategically repeated throughout the year to maintain the voltage within acceptable limits, underscoring the critical role of the AQSTSS in capturing and managing seasonal transmission system dynamics.

\subsection{Estimation of Losses Based on Time-Steps Simulation}\label{subsec:loss_estimation}
\begin{figure}[htbp]
    \centering
    \includegraphics[width=1.0\columnwidth]{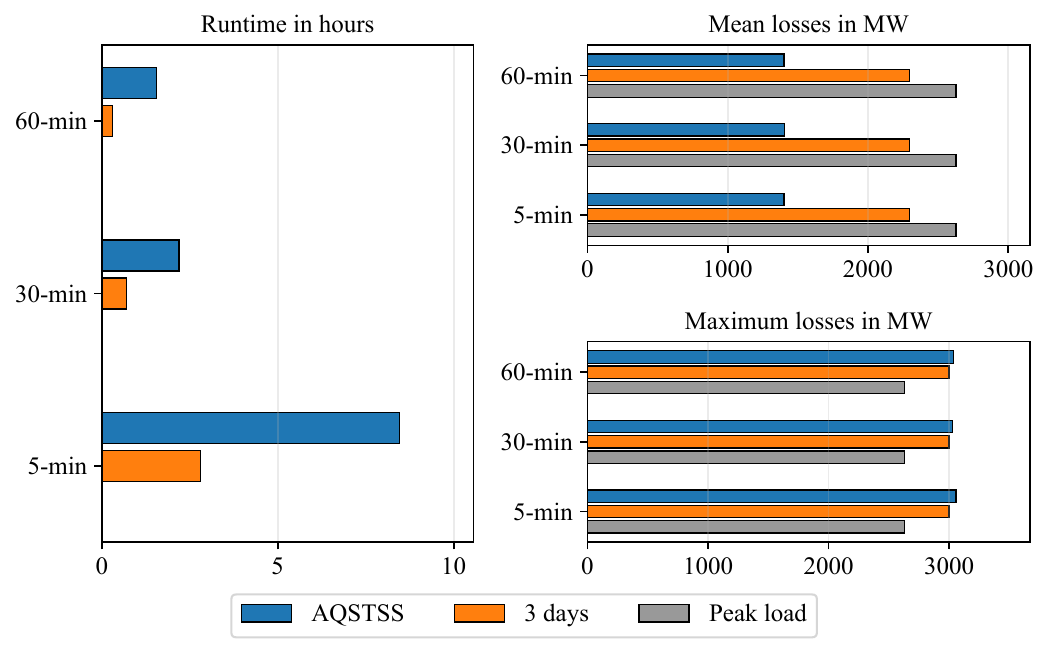}
    \caption{Simulation duration and system losses assessment}
    \label{fig:losses}
\end{figure}

Managing discrete variables and equipment control to ensure year-round network operability becomes more complex when variations in the network profiles and topology are minimal. As shown in Fig.~\ref{fig:losses}, the simulation time increases with shorter time-steps. Allowing flexibility in time-step selection improves the simulation speed tailored to the study objectives. The system losses are influenced by the network topology, voltage magnitudes, and angles across all buses. To assess these losses, the results from a single PPS were compared with simulations over three peak-load days and a full year. 
The results indicate that PPS does not capture the highest loss conditions, with maximum PPS losses approximately 500 MW lower than those observed with the chronological run. For \textbf{S1}, the average losses were approximately 50\% of the maximum observed value. Loss estimates remain consistent across time‑steps, indicating that a 60‑minute resolution is adequate for annual energy‑loss studies. 
As losses are a key criterion for siting new resources and selecting optimal network reinforcements, such deviations may influence planners toward specific reinforcement options or additional capacity.
\color{black}
\vspace{-5.5pt}
\subsection{Line Operations and Voltage Control at 735 kV}\label{subsec:vctrl}
\begin{figure}[htbp]
    \centering
    \includegraphics[width=1.0\columnwidth]{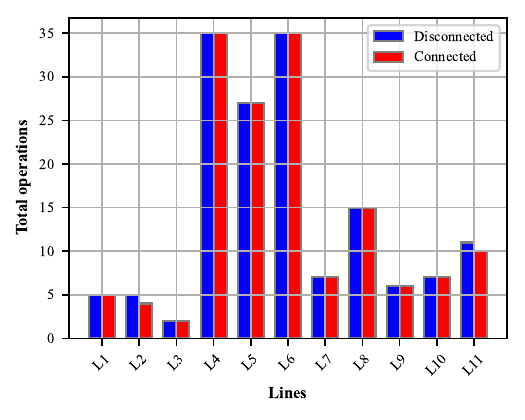}
    \caption{Status-based switching actions across strategic 735 kV lines}
    \label{fig:switching}
\end{figure}

In Hydro-Québec’s transmission system, 735 kV lines are essential for voltage control and system stability due to their inherent reactive power generation and reduced current flow for a given power transfer over long distances. Fig.~\ref{fig:switching} presents the monitoring data on switching operations across 11 strategically important transmission lines, distinguishing between disconnecting and reconnecting events. High switching frequencies on certain lines suggest operational stress or frequent network reconfigurations in response to the changing system conditions. Conversely, lines with few switching events are critically important for maintaining power flow capabilities and ensuring grid reliability. A mismatch in the switching counts over a given period indicates a difference between the initial and final system states, reflecting the decisions made by
the VO to maintain a secure system. These operational patterns highlight the challenge of managing line utilization under variable load conditions and emphasize the need for coordinated control. Such variability necessitates careful coordination to ensure system reliability and optimize the use of transmission assets. Visualization provides valuable insights into the lines subject to frequent switching, which may raise reliability concerns or highlight opportunities for optimizing control strategies. 
As a guideline, any corridor exceeding a predefined switching threshold during specific periods 
may be flagged for targeted improvements, as such activity indicates a structural constraint that may drive network investments.

\color{black}

\begin{figure}[htbp]
    \centering
    \includegraphics[width=1.0\columnwidth]{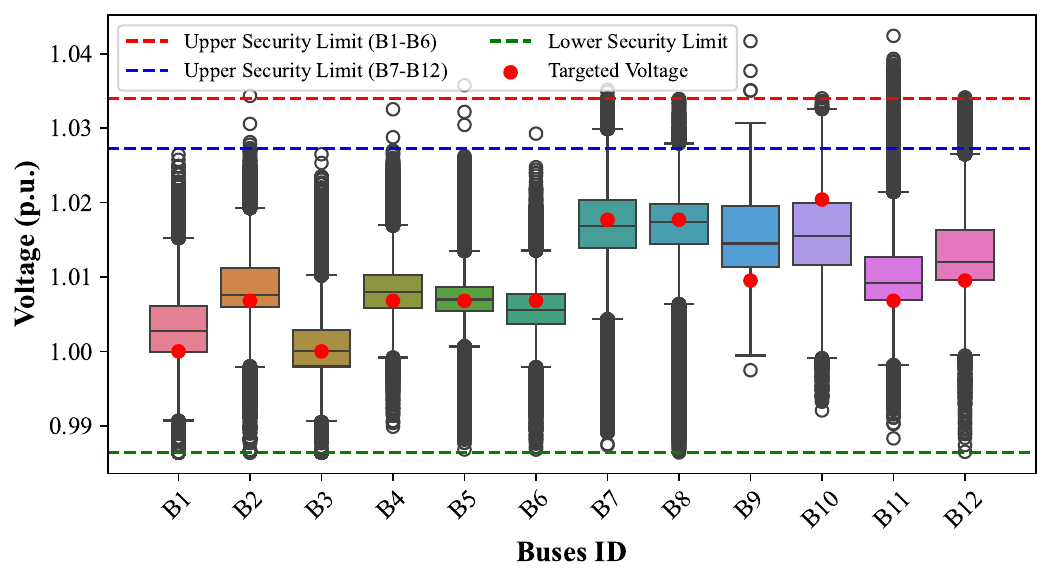}
    \caption{Statistical distribution of voltage across 735 kV main buses}
    \label{fig:boxplot}
\end{figure}
Fig.~\ref{fig:boxplot} presents a statistical analysis of voltages across the twelve 735‑kV main buses using boxplots that show the median, interquartile range, and extreme values over the full AQSTSS horizon. The dashed red and blue lines denote the upper security limits for buses B1–B6 and B7–B12, respectively, while the green line marks the lower limit; the red markers indicate the long‑term targeted voltages. Although most buses remain close to their targets, buses B5, B9, and B11 exhibit wider spreads and frequent excursions toward the security boundaries, revealing localized reactive‑power imbalances and persistent voltage‑control stress that static snapshot analyses cannot capture. These annual voltage‑stress patterns offer objective, statistically grounded indicators of where reinforcement may be required. In particular, the sustained frequency and persistence of deviations at B11 suggest that additional reactive support should be evaluated for this corridor.
\color{black}
\vspace{-4pt}
\subsection{{Impact of Wind Variability Under Critical Load Conditions}}\label{subsec:wind_variability}
To strengthen system flexibility and ensure SDB, two novel indicators were introduced: the potential increase in imports (PII) and the potential decrease in exports (PDE). These theoretical metrics quantify the available MW margin that can be leveraged to maintain SDB, offering valuable guidance for system operators, particularly in market-driven environments. Complementing these indicators, the simulations track various reserves, including synchronous, stability, and 10–30-minute reserves, to manage short-term imbalances. AGC reserves were also monitored, with corrective actions taken to maintain the threshold compliance.

\begin{figure}[htbp]
    \centering
    \includegraphics[width=1.0\columnwidth]{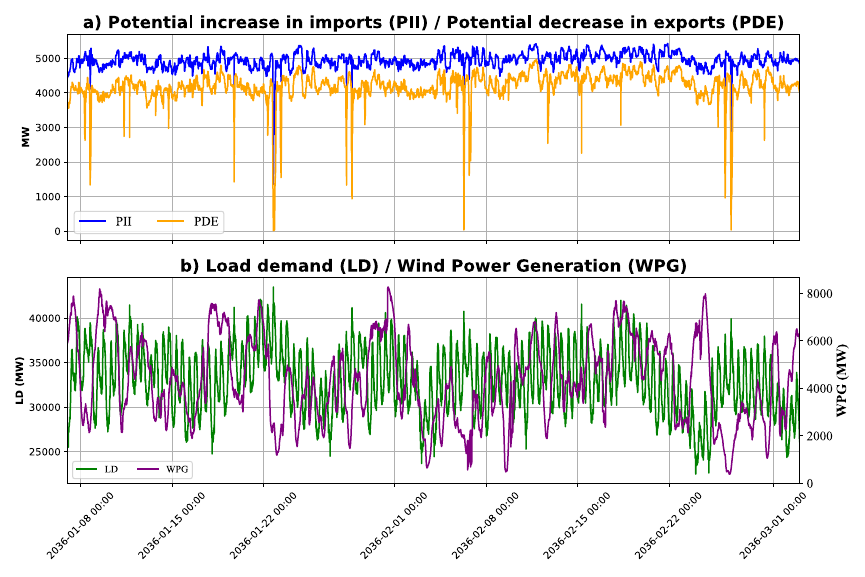}
    \caption{Monitoring emergency resources to mitigate SDB issues}
    \label{fig:monitoring}
\end{figure}

As illustrated in Fig.~\ref{fig:monitoring} a), the lowest PII and PDE values occur during periods of high demand coupled with low wind generation, precisely when the system flexibility is most constrained. This confirms the usefulness of these indicators for identifying critical operating conditions. Fig.~\ref{fig:monitoring} b) further illustrates a temporal mismatch between wind generation and system load, with wind output peaking during low-demand periods and dropping during peak demand. This dynamic is effectively captured by the AQSTSS framework and supports the development of optimized sizing and control strategies for ESSs, which are essential for mitigating renewable resources variability and ensuring grid reliability.
\color{orange}
\color{black}
\subsection{Impact of ESS on Wind Generation Variability}\label{subsec:ess_analysis}
\textbf{Scenario 2} (\textbf{S2}) extends the analysis by incorporating two BESS aggregations, totaling 1.5 GW of installed capacity, to mitigate wind generation variability. Each BESS was assigned to one of the two wind-generation zones defined in this study. 
{Table~\ref{tab:bess} summarizes the technical specifications, including power, energy capacity, and the target state-of-charge (SOC).}
\begin{table}[htbp]
    \caption{BESS Parameters for \textbf{Scenario 2}}
    \centering
        \begin{tabular}{c c c c c}
            {}& \textbf{Location} & \makecell{\textbf{Power}\\ \textbf{Capacity}\\ \textbf{(MW)}} & \makecell{\textbf{Energy}\\ \textbf{Capacity}\\ \textbf{(MWh)}} & \makecell{\textbf{Balance}\\ \textbf{SOC}\\ \textbf{(\%)}}\\
            \hline
            \rowcolor{blue!10}
            \textbf{BESS 1} & East zone & 860 & 3440 & 50\\
            \textbf{BESS 2} & Non-East zone & 640 & 2560 & 50\\
            \hline
        \end{tabular}
        \label{tab:bess}
\end{table}

\begin{figure}[htbp]
    \centering
    \includegraphics[width=1.0\columnwidth]{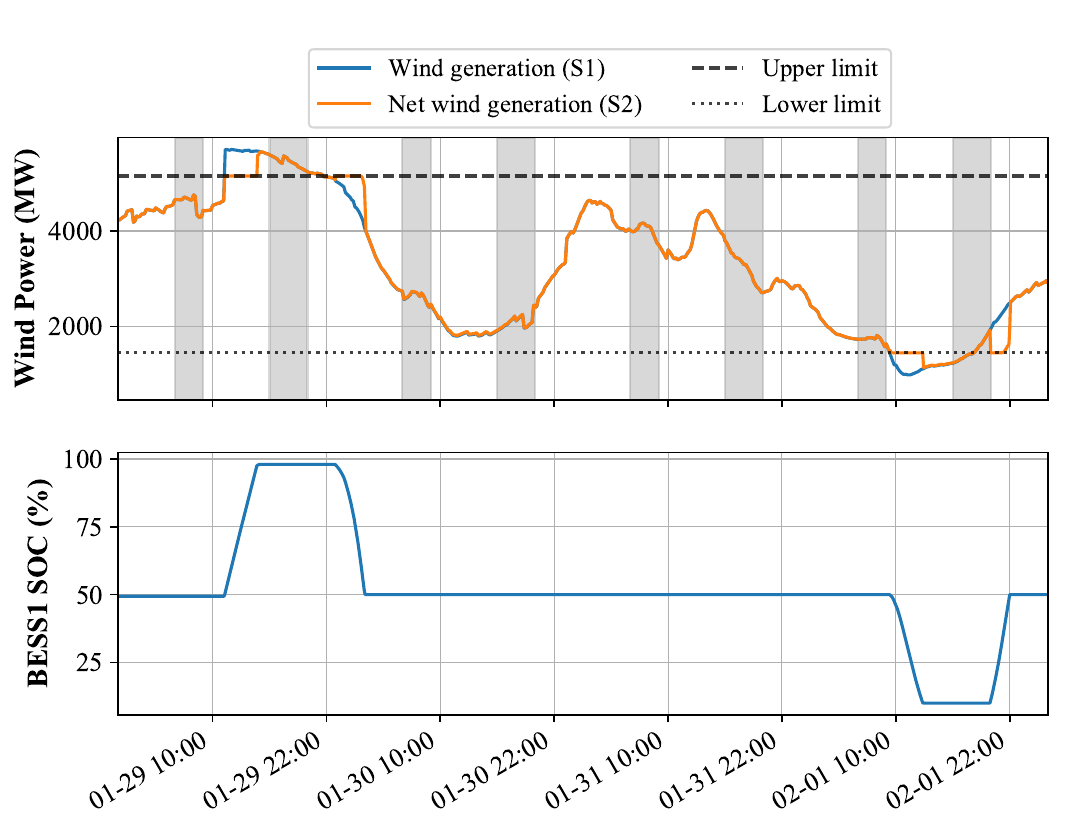}
    \caption{Wind farm generation (top) and battery storage SOC (bottom) in the East zone}
    \label{fig:wind}
\end{figure}

The control strategy outlined in Algorithm \ref{alg:ess_mode} was applied independently to each BESS aggregation, ensuring localized compensation for wind generation variability within their respective zones. The sample results from the 5-minute simulation of \textbf{S2} are presented in Fig.~\ref{fig:wind} to illustrate the behavior of the control algorithm. Over a period of three winter days, the impact of BESS 1 on the net wind farm generation and the evolution of the SOC in the East zone was observed. As expected, BESS 1 charges when the wind generation exceeds the upper limit, and discharges when the lower generation limit is exceeded. In addition, the SOC is controlled to return to a value of 50\% when the generation lies within the limits. In either case, the BESS is not allowed to charge during the peak-load periods, as indicated by the greyed-out areas. 
{Finally, it should be noted that in this study, the power and energy capacities as well as the balance SOC of 50\% were used as baseline values for these parameters. This choice was made to illustrate a typical workflow for a user of the proposed AQSTSS framework. In this case, the TSP wishes to study the impacts of energy storage in two predetermined locations, but they may not know how to size and parametrize the BESSs. Further analyses, presented in this section, demonstrate how the simulation results can be interpreted by the TSP to better design the energy storage parameters.}

\begin{figure}[htbp]
    \centering
    \includegraphics[width=1.0\columnwidth]{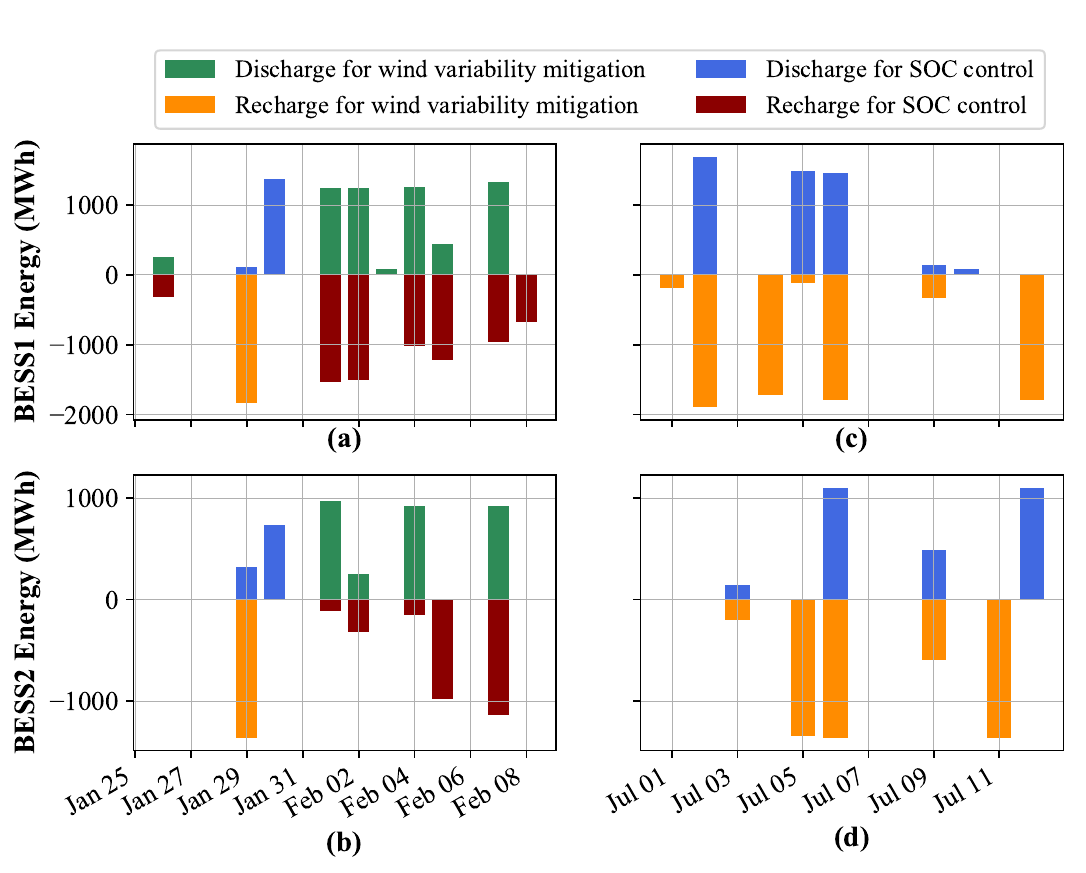}
    \caption{Daily BESS energy charged and discharged during two peak load weeks in (a) the east zone, (b) the non-east-zone, and during two low load weeks in (c) the East zone and in (d) the non-East zone}
    \label{fig:ess_energy}
\end{figure}

A visual inspection of the net wind generation curve in Fig.~\ref{fig:wind} revealed that the energy capacity of BESS 1 was insufficient to fully mitigate prolonged periods of over- or under-generation. This issue could be addressed by investing in long-duration energy storage, such as 8- or 12-hour battery systems. However, while such a solution may be adequate for the limited time frame of the results in Fig.~\ref{fig:wind}, it might not be optimal in the context of annual TSEP. Therefore, the proposed AQSTSS framework was utilized to conduct a more extensive analysis of the installed BESSs over one year. In Fig.~\ref{fig:ess_energy}, the daily charged and discharged energies of both BESSs during different periods are analyzed and compared. Specifically, two peak‑load winter weeks and two low‑load summer weeks were analyzed.

Color coding was used to distinguish the energy charged or discharged for wind variability mitigation from that used for SOC regulation. Nonetheless, during the two peak-load weeks, wind generation was generally lower, necessitating the discharge of BESS units. Conversely, the relatively higher wind generation during the two low-load weeks enabled BESSs to primarily absorb excess energy. 
{This seasonal behavior suggests that maintaining a fixed SOC balance of 50\% may not be adequate for year-round operation. In practice, the SOC balance is a simulation parameter that is only set once at the start of the AQSTSS. Therefore, the charging/discharging behavior of the BESS, determined by the control strategy in Algorithm \ref{alg:ess_mode}, remains agnostic to present and future values of load and wind generation profiles during the simulation.} 

\begin{figure}[htbp]
    \centering
    \includegraphics[width=1.0\columnwidth]{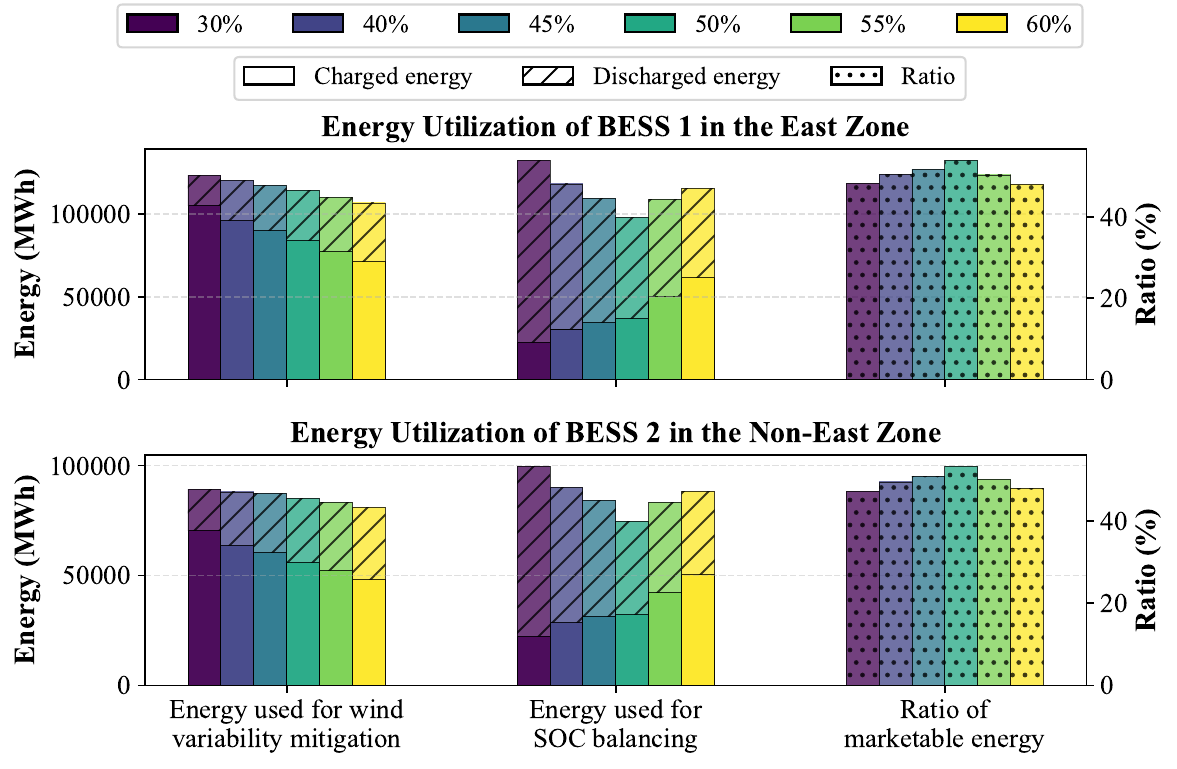}
    \caption{Comparison of the annual energy utilizations of BESS 1 (top) and BESS 2 (bottom) for the different SOC balance cases}
    \label{fig:hist_soc}
\end{figure}

{A sensitivity analysis on the value of the SOC balance was performed to illustrate the effect of this control parameter on the utilization of energy storage. Including the baseline case with 50\% SOC balance, \textbf{S2} was simulated with six different SOC balance values as shown in Fig.~\ref{fig:hist_soc}.}

{The plots depict yearly energy utilization which includes both charged and discharged energy. This utilization is classified between energy used for wind generation variability mitigation and energy used for SOC balancing. In addition, the ratio of marketable energy represents the percentage of the total energy that was used to mitigate the variability of wind generation. We can first observe that energy used for wind variability mitigation decreases as the SOC balance value increases. This result implies that the BESSs need to recharge to compensate for over-generation more often than they need to discharge to mitigate under-generation over the year.
In contrast, the lowest amount of energy used for SOC balancing occurs at the 50\% case along with increasing energy utilization when the SOC balance deviates from its baseline value. In other words, the SOC control effort increases as the SOC balance value moves further away from 50\%. As a result, the highest ratio of marketable energy also occurs at the 50\% case. Depending on their specific expansion planning objectives and criteria, TSPs may use these results to make more informed decisions when designing the SOC control parameters. For instance, if the goal is purely to maximize the utilization of BESSs for wind variability mitigation, then a lower SOC balance like 30\% would be more appropriate. Conversely, keeping the SOC balance value closer to 50\% would reduce the absolute amount of marketable energy, but would also allow the BESSs to offer additional services that may add further value to the system (e.g. reserves).}
\begin{figure}[htbp]
    \centering
    \includegraphics[width=1.0\columnwidth]{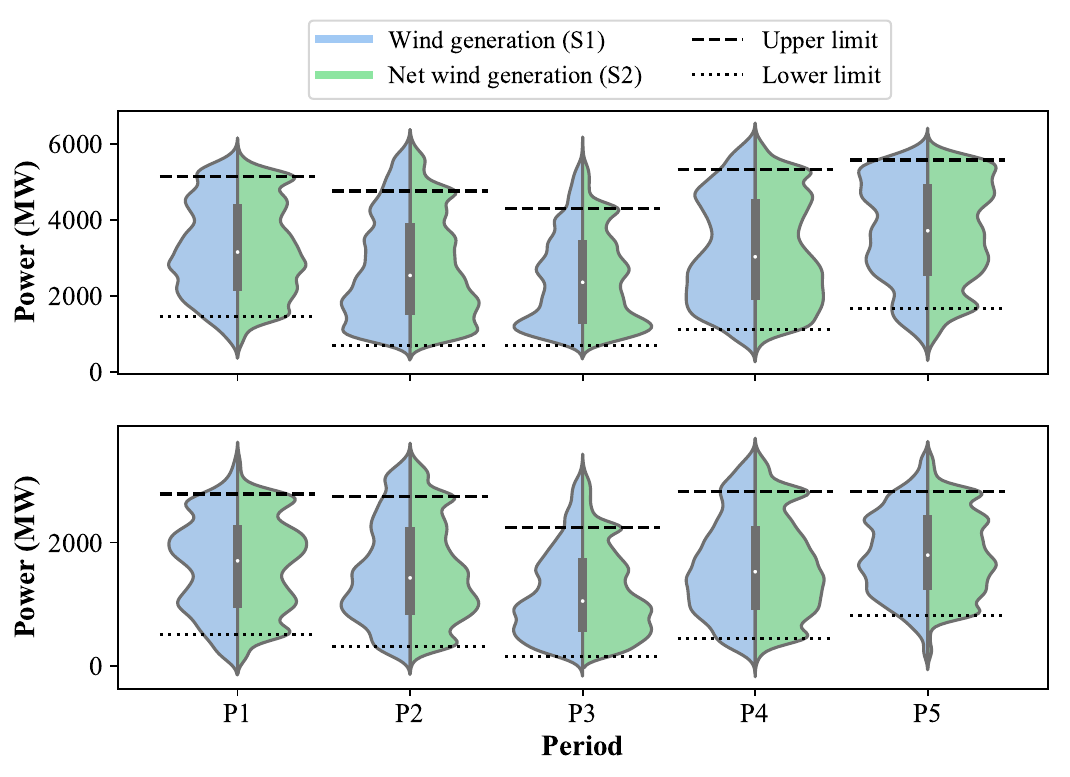}
    \caption{Distributions of wind farm generation in the East zone (top) and in the non-East zone (bottom)}
    \label{fig:violons}
    \vspace{-4pt}
\end{figure}
\begin{table*}[!b]
    \caption{{Translating AQSTSS Outputs Into Planning Decisions}}
    \label{tab:aqstss_to_decisions}
    \centering
    \renewcommand{\arraystretch}{1.}    
    \begin{tabular}{%
        p{3.cm}  
        p{4.cm}  
        p{9.8cm}  
    }
        \hline
        \centering\arraybackslash\textbf{{Simulation metric}} &
        \centering\arraybackslash{\textbf{{Interpretation for TSPs}}} &
        \centering\arraybackslash{\textbf{{Actionable decision-making criteria}}} \\
        \hline
        \rowcolor{blue!10}
        Recurring thermal overload hours on key corridors &
        Identifies structurally congested interfaces and off-peak bottlenecks not captured by snapshot methods. &
        \begin{minipage}[t]{\linewidth}
         \begin{itemize}
             \item Add new lines where overloads exceed a hours/year threshold.
            \item Reconfigure regional transfers.
            \item Evaluate non-wire solutions if overloads are seasonal and short-duration.
         \end{itemize}
         \end{minipage}         
        \\
       Seasonal voltage excursions &
       Reveals voltage control deficiencies under different seasonal operating modes. &
         \begin{minipage}[t]{\linewidth}
         \begin{itemize}
             \item Install new reactive resources; adjust tap-changer settings.
            \item Install VAR compensators at buses exceeding the \textit{N}‑event voltage‑deviation threshold per year.
         \end{itemize}
         \end{minipage} \\
         \rowcolor{blue!10}
       Discrete switching actions &
       Indicates stress on equipment from rising VRE/DER variability.
       &
        \begin{minipage}[t]{\linewidth}
         \begin{itemize}
             \item Deploy ESS/DR to reduce excessive tap operations.
            \item Upgrade on-load tap-changer transformers approaching operational limits.
            \item Provision additional maintenance and monitoring actions to control risk.
         \end{itemize}
         \end{minipage}            
        \\
        Wind curtailment patterns &
         Highlights structural energy spillage due to network limits. &
         \begin{minipage}[t]{\linewidth}
         \begin{itemize}
                      \item Add new transmission equipment when curtailment exceeds an economic threshold.
            \item Justify the economic value of ESSs through their ability to reduce curtailment.
         \end{itemize}
         \end{minipage} \\
           \rowcolor{blue!10}
        ESS dispatch effectiveness &
        Shows whether ESSs can defer or avoid traditional reinforcements. &
        \begin{minipage}[t]{\linewidth}
         \begin{itemize}
         \item Defer transmission upgrades if ESSs mitigates events.
            \item Prioritize ESS siting where service stacking opportunities exist, including voltage support, ramping, and other ancillary services. 
         \end{itemize}
         \end{minipage}          
        \\ 
         Reserve margin violations 
         &
        Reveals when traditional planning underestimates balancing needs.
        &
         \begin{minipage}[t]{\linewidth}
         \begin{itemize}
             \item Justify new flexible resources, including interconnection contract.
         \end{itemize}
         \end{minipage} \\
         \rowcolor{blue!10}
       Critical off‑peak cases  &
       Detects low-inertia conditions, voltage or thermal issues missed by peak‑only studies.
       &
        \begin{minipage}[t]{\linewidth}
         \begin{itemize}
                      \item Include off‑peak dimensioning cases in reinforcement studies.
            \item Reconfigure seasonal transmission topology.
         \end{itemize}
         \end{minipage}            
        \\
        \hline
    \end{tabular}
\end{table*}
\color{black}

{Furthermore, the AQSTSS results may serve as guide for the user to further fine-tune the SOC balance values for different time periods. For example, the results can be used to identify the periods during which energy storage is primarily solicited for charging or discharging, as illustrated Fig.~\ref{fig:ess_energy}. When extended to the entire year, this analysis would allow the user to set higher SOC balance values during periods of high demand for discharging, and lower values during periods of high demand for charging. This adjustment may help improve the availability of BESS energy for subsequent studies, thus reducing instances when the BESS energy is insufficient to compensate for wind variability over extended periods.}

Using the AQSTSS tool, detailed statistics on the impact of BESSs on wind generation variability can be extracted over the entire simulation year. This allowed us to identify temporal trends and seasonal performance patterns. Indeed, Fig.~\ref{fig:violons} illustrates the distribution of wind power generation in both zones for each of the five seasonal periods $p \in\{1,2,3,4,5\}$. 
Additionally, it highlights the upper and lower generation limits for each period and compares base wind power in \textbf{S1} with net wind power including BESS in \textbf{S2}.

From these graphs, the effectiveness of the BESSs in mitigating wind variability can be assessed across seasons. For instance, during P1 in the East zone, the BESS successfully charged and shifted a significant amount of wind power generation below the upper limit. Conversely, during P5, the BESS had a limited impact on the net generation. This behavior may suggest that the BESS reaches its full energy capacity more frequently in P5 than in P1, thus preventing necessary charging. Alternatively, it could also indicate that wind generation does not often exceed the upper limit, making the effect of storage less visible in the distribution. 

To ascertain the actual cause of this behavior, more granular data from P5 must be analyzed. The AQSTSS tool facilitates this by providing the relevant wind and energy storage results at a 5-minute resolution, as demonstrated in Fig.~\ref{fig:wind}. Extending this analysis across all periods and both generation limits allows for more informed decisions regarding BESS sizing and control strategies tailored to specific use cases.


These insights are particularly valuable for transmission system planners as energy storage applications grow in scale and complexity. By enabling year‑long, high‑resolution evaluation of ESS performance, the AQSTSS framework clarifies how storage parameters and control strategies affect system behavior, asset utilization, and grid reliability.
\subsection{Decision-Making Insights Drawn from AQSTSS Results}
As a non‑exhaustive overview, Table~\ref{tab:aqstss_to_decisions} highlights how key AQSTSS outputs map to actionable planning criteria, thereby supporting both wire and non‑wire reinforcement strategies.
By linking each metric to the underlying operational challenge, the table clarifies how AQSTSS distinguishes between persistent and seasonal constraints, enables targeted deployment of flexibility resources, and captures off‑peak conditions typically overlooked in snapshot analyses. 
\color{black}
\section{Conclusions}\label{sec:conclusion}
This paper introduces an annual quasi-static time-series simulation framework for transmission system expansion planning, addressing the increasing complexity caused by the high penetration of distributed energy resources and variable renewable generation. The proposed methodology enables year-round, high-resolution modeling of system behavior, capturing not only operational constraints and voltage control dynamics but also the time-dependent states of transmission system equipment. This temporal granularity allows planners to evaluate equipment utilization, switching actions, and control strategies under realistic and evolving conditions.

Applied to Hydro-Québec’s projected 2035/2036 transmission system, the AQSTSS framework provides critical insights into operational challenges and flexibility opportunities. The case studies demonstrated the impact of doubling the wind power capacity, tenfold increase in EV-related demand, and integration of 1.5 GW energy storage systems to mitigate renewable variability 
{and to support the supply–demand balance across seasonal and daily variations.}

Overall, AQSTSS offers a robust and scalable foundation for aligning long-term planning with operational realities, enabling a more resilient, efficient, and adaptive transmission system development in the evolving energy landscape.

 
%
\bibliographystyle{IEEEtran}
\bibliography{references_aqstss.bib}

 


\vspace{-20pt} 
\begin{IEEEbiographynophoto}{Hussein Suprême}
received the B.Sc. degree in electromechanical engineering from Université d’État d’Haïti in 2007, and the M.Sc.A. and Ph.D. degrees from École de technologie supérieure, Canada, in 2013 and 2017, respectively.

He is a Researcher at the Institut de Recherche d’Hydro-Québec (IREQ), where he focuses on the integration and impact assessment of distributed energy resources, power system dynamics and control, and energy system optimization. He also leads R\&D initiatives in future grid planning and operations.

Dr. Suprême is an Associate Professor at Polytechnique Montréal and serves as an international consultant for renewable energy integration and power system stability.
\end{IEEEbiographynophoto}
\vspace{-33pt}
\begin{IEEEbiographynophoto}{Martin de Montigny}
received the B.Eng., M.Eng., and Ph.D. degrees in electrical engineering from the Université du Québec à Trois-Rivières in 1997, 2000 and 2006, respectively.

Since 2007, he has been working as a Researcher with the Network Control and Management Department, Hydro-Québec Research Institute. He specializes in evaluating the transmission grid impacts of variable renewable and distributed energy resources.

Dr. de Montigny is the Project Manager of OSER initiative, which aims to design and implement new tools for the Transmission planners. 
\end{IEEEbiographynophoto}
\vspace{-33pt}
\begin{IEEEbiographynophoto}{Kevin-R. Sorto-Ventura}
received the B.Eng. degree in electrical engineering from McGill University, Montreal, QC, Canada in 2019, and the M.A.Sc. degree in electrical engineering from École de technologie supérieure, Montreal, QC, Canada in 2021.

He is currently a researcher at the Institut de Recherche d’Hydro-Québec (IREQ), as well as a Ph.D. student at McGill University. His main research interests are the modeling and simulation of the impacts of inverter-based resources in the context of transmission system planning. 
\end{IEEEbiographynophoto}
\vspace{-35pt}
\begin{IEEEbiographynophoto}{Hind Chit Dirani}
received the electrical engineering diploma degree from the Lebanese University in 2008, the M.Sc. degree in Renewable Energy from the Lebanese University and Saint Joseph University, Lebanon, in 2013, and the Ph.D. degree in Electrical Engineering from the École de Technologie Supérieure, Montreal, QC, Canada, in 2018. She is currently a Researcher at the Hydro-Québec Research Institute (IREQ), Varennes, QC, Canada. 
\end{IEEEbiographynophoto}
\vspace{-35pt}
\begin{IEEEbiographynophoto}{Mouhamadou Makhtar Dione}
received the M.Eng. degree in electrical engineering from ENSIEG/INPG, Grenoble, France, in 2000.

He has been with Hydro-Québec’s Research Institute since 2011, where he is a Researcher in transmission and distribution planning in the context of high distributed energy resources integration. 
\end{IEEEbiographynophoto}
\vspace{-35pt}
\begin{IEEEbiographynophoto}{Nicolas Compas}
received the B.Eng. degrees in electrical engineering from Polytechnique Montreal, Canada in 2007. Since, he has been working as a transmission system planner at Hydro-Québec. He is an expert in the modeling, simulation, and analysis of the dynamic behavior of transmission systems, with a specialization in emerging technologies such as high-voltage direct current (HVDC) solutions and distributed energy resources (DER). He is also an active member of various working groups within the North American Electric Reliability Corporation (NERC).
\end{IEEEbiographynophoto}

\vfill

\end{document}